\documentstyle[aps,epsfig,prl]{revtex}
\begin{document}
\title{Observation of the Decay
$K_L\rightarrow \mu^+\mu^- \gamma \gamma$.}
\newdimen\fullhsize\newdimen\play\newdimen\hmargins\newdimen\vmargins
\def\newmargins#1#2{\hmargins=#1\vmargins=#2
   \play=8.5truein\advance\play by-2\hmargins\global\fullhsize=\play
   \advance\play by-4\hmargins\divide\play by3\global\hsize=\play
   \play=11truein\advance\play by-2\vmargins\global\vsize=\play
   \play=-1truein\advance\play by\hmargins\global\hoffset=\play
   \play=-0.65truein\advance\play by\vmargins\global\voffset=\play}
\newcommand {\klpmz} {$K_L\rightarrow \pi^+\pi^-\pi^0$}
\author{
A.~Alavi-Harati$^{12}$,
I.F.~Albuquerque$^{10}$,
T.~Alexopoulos$^{12}$,
M.~Arenton$^{11}$,
K.~Arisaka$^2$,
S.~Averitte$^{10}$,
A.R.~Barker$^5$,
L.~Bellantoni$^7$,
A.~Bellavance$^9$,
J.~Belz$^{10}$,
R.~Ben-David$^7$,
D.R.~Bergman$^{10}$,
E.~Blucher$^4$, 
G.J.~Bock$^7$,
C.~Bown$^4$, 
S.~Bright$^4$,
E.~Cheu$^1$,
S.~Childress$^7$,
R.~Coleman$^7$,
M.D.~Corcoran$^9$,
G.~Corti$^{11}$, 
B.~Cox$^{11}$,
M.B.~Crisler$^7$,
A.R.~Erwin$^{12}$,
R.~Ford$^7$,
A.~Glazov$^4$,
A.~Golossanov$^{11}$,
G.~Graham$^4$, 
J.~Graham$^4$,
K.~Hagan$^{11}$,
E.~Halkiadakis$^{10}$,
K.~Hanagaki$^8$,  
M.~Hazumi$^8$,
S.~Hidaka$^8$,
Y.B.~Hsiung$^7$,
V.~Jejer$^{11}$,
J.~Jennings$^2$,
D.A.~Jensen$^7$,
R.~Kessler$^4$,
H.G.E.~Kobrak$^{3}$,
J.~LaDue$^5$,
A.~Lath$^{10 \dagger}$,
A.~Ledovskoy$^{11}$,
P.L.~McBride$^7$,
A.P.~McManus$^{11}$,
P.~Mikelsons$^5$,
E.~Monnier$^{4,*}$,
T.~Nakaya$^7$,
K.S.~Nelson$^{11}$,
H.~Nguyen$^7$,
V.~O'Dell$^7$, 
M.~Pang$^7$, 
R.~Pordes$^7$,
V.~Prasad$^4$, 
C.~Qiao$^4$, 
B.~Quinn$^4$,
E.J.~Ramberg$^7$, 
R.E.~Ray$^7$,
A.~Roodman$^4$, 
M.~Sadamoto$^8$, 
S.~Schnetzer$^{10}$,
K.~Senyo$^8$, 
P.~Shanahan$^7$,
P.S.~Shawhan$^4$,
W.~Slater$^2$,
N.~Solomey$^4$,
S.V.~Somalwar$^{10}$, 
R.L.~Stone$^{10}$, 
I.~Suzuki$^8$,
E.C.~Swallow$^{4,6}$,
R.A.~Swanson$^{3}$,
S.A.~Taegar$^1$,
R.J.~Tesarek$^{10}$, 
G.B.~Thomson$^{10}$,
P.A.~Toale$^5$,
A.~Tripathi$^2$,
R.~Tschirhart$^7$, 
Y.W.~Wah$^4$,
J.~Wang$^1$,
H.B.~White$^7$, 
J.~Whitmore$^7$,
B.~Winstein$^4$, 
R.~Winston$^4$, 
T.~Yamanaka$^8$,
E.D.~Zimmerman$^4$}
\address{
$^1$ University of Arizona, Tucson, Arizona 85721 \\
$^2$ University of California at Los Angeles, Los Angeles, California 90095 \\
$^{3}$ University of California at San Diego, La Jolla, California 92093 \\
$^4$ The Enrico Fermi Institute, The University of Chicago, 
Chicago, Illinois 60637 \\
$^5$ University of Colorado, Boulder, Colorado 80309 \\
$^6$ Elmhurst College, Elmhurst, Illinois 60126 \\
$^7$ Fermi National Accelerator Laboratory, Batavia, Illinois 60510 \\
$^8$ Osaka University, Toyonaka, Osaka 560 Japan \\
$^9$ Rice University, Houston, Texas 77005 \\
$^{10}$ Rutgers University, Piscataway, New Jersey 08855 \\
$^{11}$ The Department of Physics and Institute of Nuclear and 
Particle Physics, University of Virginia, 
Charlottesville, Virginia 22901 \\
$^{12}$ University of Wisconsin, Madison, Wisconsin 53706 \\
$^{*}$ On leave from C.P.P. Marseille/C.N.R.S., France \\
$^{\dagger}$ To whom correspondence should be addressed. \\
}
\maketitle
\begin{abstract}
We have observed the decay $K_L\rightarrow \mu^+\mu^- \gamma \gamma$
at the KTeV  experiment at Fermilab.     This decay presents a 
formidable background to the search for new physics in
$K_L\rightarrow\pi^0\mu^+\mu^-$.
The 1997 data yielded
a sample of 4 signal events, with an expected 
background of 0.155 $\pm$ 0.081 events. The 
branching ratio is  
${\mathcal B}$($K_L\rightarrow \mu^+\mu^- \gamma \gamma$) $ =
(10.4^{+7.5}_{-5.9}  {\rm ~(stat)} \pm 0.7 {\rm ~(sys)})\times 10^{-9}$ with 
$m_{\gamma\gamma} \geq 1 {\rm ~MeV/c}^2$, consistent with
a QED calculation which predicts $(9.1\pm 0.8)\times 10^{-9}$.
\end{abstract}

In this paper we present the first measurement of the
branching ratio for $K_L\rightarrow\mu^+\mu^-\gamma\gamma$.
This decay  is expected to proceed 
mainly via the Dalitz decay $K_L\rightarrow\mu^+\mu^-\gamma$
with an internal bremsstrahlung photon.   
This decay is one of a family of radiative decays 
($K_L\rightarrow\mu^+\mu^-\gamma$, $K_L\rightarrow\mu^+\mu^-\gamma\gamma$,
$K_L\rightarrow e^+ e^-\gamma$, $K_L\rightarrow e^+ e^-\gamma\gamma$)
which are under study at KTeV and elsewhere \cite{ref:raddecays,ref:mmg}.
The decay   
$K_L\rightarrow\mu^+\mu^-\gamma\gamma$ 
presents a formidable background to the search
for direct CP 
violation and new physics 
in $K_L\rightarrow \pi^0\mu^+\mu^-$ decays \cite{ref:pimumu}.


The measurement presented here was performed as part of the KTeV 
experiment, which has been described elsewhere~\cite{ref:ktev}.
The experiment
used two nearly parallel  $K_L$ beams
created by 800 GeV protons incident on a BeO target.  The  decays
used in our studies were collected in a  region approximately
65 meters long, situated 94 meters from the 
production target.  The fiducial volume was surrounded by a photon veto
system used to reject events in which photons missed the calorimeter.
The charged particles were detected by four
drift chambers, each consisting of one horizontal and one vertical
pair of planes, with typical resolution of 70 $\mu m$ per plane pair.
Two drift chambers were situated on either side of an analysis magnet
which  imparted  205 MeV/c of transverse momentum to
charged particles.  The drift chambers were followed by a
trigger hodoscope bank, and a 3100 element pure CsI calorimeter
with  electromagnetic 
energy resolution of 
$\sigma(E)/E = 0.45\% \oplus 2.0\%/\sqrt{E{\rm (GeV)}} $.  
The calorimeter was followed by a
muon filter composed of 
a 10 cm thick lead wall and three steel walls
totalling 511 cm.  
Two planes of scintillators situated
after the third steel wall served to identify muons.  
The planes had 15 cm segmentation, one horizontal, the other
vertical.
 
The trigger 
for the signal events 
required  hits in 
the upstream  drift chambers
consistent with two tracks,  as well as two hits in the trigger 
hodoscopes.  The calorimeter was required to have at least one
cluster with over 1 GeV in energy, within a narrow (20 ns) time gate. 
The  muon
counters were required to have at least two hits.
In addition, preliminary online identification of these decays 
required reconstruction of two track candidates originating from
a loosely-defined vertex, and each of those track candidates was
required to point to a cluster
in the calorimeter with energy less than $ 5 {\rm ~GeV}$.
A separate trigger was used to collect $K_L\rightarrow\pi^+\pi^-\pi^0$
decays which were used for normalization.  This trigger was similar
to the signal trigger but had no requirements on 
hits in the muon hodoscopes or clusters in the calorimeter.  The 
preliminary online identification was performed on the normalization
sample as well, but no energy requirements were made on clusters
pointed to by the tracks.  The normalization mode trigger was
prescaled by a factor of 500:1.

The main background to $K_L\rightarrow\mu^+\mu^-\gamma\gamma$
was the Dalitz decay $K_L\rightarrow\mu^+\mu^-\gamma$ with
an additional cluster in the calorimeter 
coincident with  but not due to the
decay.  Such an ``accidental'' cluster could appear as a photon.  
Additional backgrounds were 
$K_L\rightarrow\pi^+\pi^-\pi^0$ decays with the 
charged pions misidentified as muons 
due to pion decay or pion punchthrough the filter steel, and 
$K_L\rightarrow \pi^{\pm} \mu^{\mp} \nu$ decays ($K_{\mu3}$)
with both charged pion misidentification and accidental cluster contributions.
Other contributions, such as $K_L\rightarrow\pi^+\pi^-$ decays
and $K_L\rightarrow\pi^+\pi^-\gamma$ decays, were negligible.

Offline analysis of the signal 
required the full reconstruction of exactly two tracks.
The vertex reconstructed from the two tracks 
was required to fall 
between 100 meters and 158 meters from the target.
In order to reduce backgrounds due to pion decay in flight,
we required that the track segments upstream and downstream
of the analysis magnet matched  
to within 1 mm  at the magnet bend plane.
Further, we required the  $\chi^2$
calculated from the reconstructed two-track vertex  be less than 
10 for 1 degree of freedom.
Tracks were required to have  momenta equal to or greater than 
10  GeV/c 
to put them above threshold  for passing through the filter steel
but below 100 GeV/c to ensure well measured track momenta.
Since muons typically deposit $\sim$ 400 MeV in the calorimeter,
we required the energy deposited by each track be 1 GeV or less.  
In addition, we required 
two non-adjacent hits in both the vertically
and horizontally segmented muon counters.

\begin{figure}
\centerline{\epsfysize 3.0 in \epsfbox{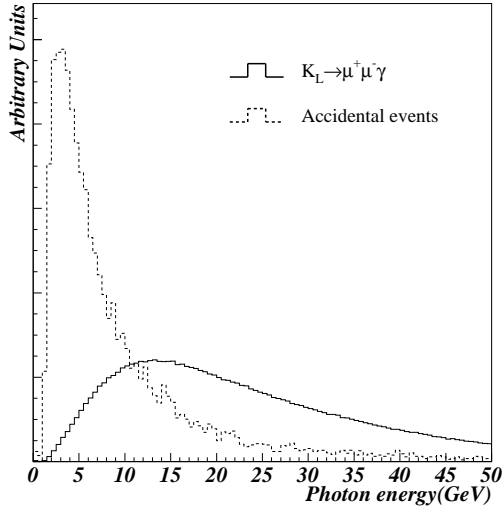}}   
\caption[]{Energy deposited in the calorimeter by photons from
Monte Carlo simulations of 
$K_L\rightarrow\mu^+\mu^-\gamma$ events (solid) vs. accidental
clusters (dashed) from data taken with a random trigger.  
}
\label{fig:hccacc}
\end{figure}

Figure~\ref{fig:hccacc} shows the expected distribution of 
cluster energy due to photons from $K_L\rightarrow\mu^+\mu^-\gamma$
events and those from accidental sources.
Accidental clusters in the calorimeter were  typically of 
low energy. 
Events were required to have two calorimeter clusters 
consistent with 
photons with no tracks pointing to them. One of these clusters 
was required
to have greater than 10 GeV of energy, thus reducing backgrounds due to
accidental clusters.  

In order to reject backgrounds from decays that contained a $\pi^0$,
the invariant mass of the two photons, $m_{\gamma\gamma}$, was
required to be less than 130 MeV/$c^2$.
Approximately 8\% of the  $K_L\rightarrow \pi^+\pi^-\pi^0$ 
decays in which  the
charged pions decay to muons survived the $m_{\gamma\gamma}$ cut because
the mismeasurement of the charged vertex
smeared the  $m_{\gamma\gamma}$ distribution.  In order to remove these
events, we constructed a variable 
(${\mathrm R}^{\pi\pi}_{\parallel}$) defined as  
\begin{equation}
{{\mathrm R}^{\pi\pi}_{\parallel}} = 
\frac{(m_K^2 - m_{\pi\pi}^2 - m_{\pi^0}^2)^2 
- 4m_{\pi\pi}^2 m_{\pi^0}^2 -  4m_K^2 p_{\perp \pi\pi}^2}
{p_{\perp \pi\pi}^2 + m_{\pi\pi}^2}
\end{equation}
where $m_K$ is the kaon mass, $m_{\pi\pi}$ is the invariant mass of the
two tracks assuming they are due to charged pions,  
$p_{\perp \pi\pi}^2$ is the square of the transverse momentum
of the two pions with respect to a line connecting the target to the
two-track vertex, and  $m_{\pi^0}$ is the mass of the $\pi^0$.
This quantity is proportional to 
the square of the longitudinal momentum of the $\pi^0$ 
in a frame along the $K_L$ flight direction
where the $\pi^+\pi^-$ pair has no longitudinal momentum.

\begin{figure}
\centerline{\epsfysize 2.5 in \epsfbox{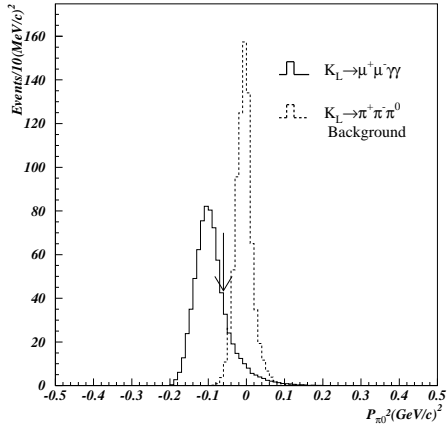}}   
\caption[]{The distribution of ${\mathrm R}^{\pi\pi}_{\parallel}$
(see text) 
from Monte Carlo simulations of the signal (solid) and
backgrounds from $K_L\rightarrow\pi^+\pi^-\pi^0$ (dashed)
that remain after the $m_{\gamma\gamma}<130 {\rm MeV/c}^2$ 
requirement. The arrow
indicates the cut at -0.06, above which events were discarded.}
\label{fig:pp0kin}
\end{figure}

Figure~\ref{fig:pp0kin}
shows the expected distribution of 
${\mathrm R}^{\pi\pi}_{\parallel}$ for the signal
(using an $\mathcal O(\alpha)$ QED matrix element),  and  the
$K_L\rightarrow \pi^+\pi^-\pi^0$ background.
By requiring  
${\mathrm R}^{\pi\pi}_{\parallel}$ to be -0.06 or less, 
92.7\% of the remaining
$K_L\rightarrow \pi^+\pi^-\pi^0$  background was eliminated.

\begin{figure}
\centerline{\epsfysize 3.0 in \epsfbox{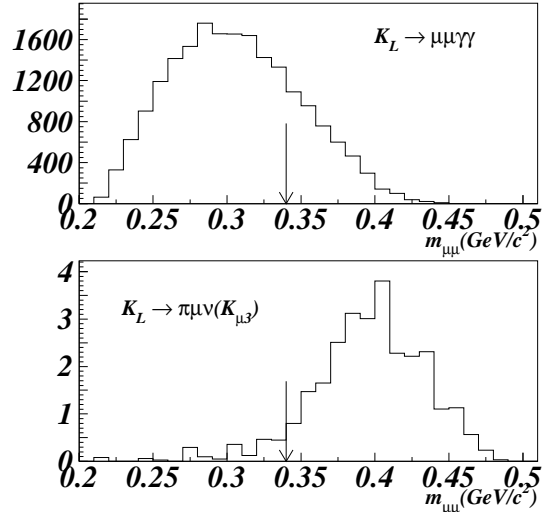}}   
\caption[]{The distribution of $m_{\mu\mu}$ from 
Monte Carlo simulations of the signal (top) and
backgrounds from $K_{\mu 3}$ (bottom).  The arrows indicate the cut
at 340 MeV/$c^2$, above which events were discarded.}
\label{fig:mmumu}
\end{figure}

The invariant mass
of the two tracks assuming muons, 
$m_{\mu\mu}$, provided a way
to reduce backgrounds due to $K_{\mu 3}$ decays.
Figure~\ref{fig:mmumu} shows the expected distribution
of $m_{\mu\mu}$ for the signal and background. 
We required $m_{\mu\mu}$ to be 
less than 340 MeV/$c^2$.
This cut eliminated
92.9\% of the $K_{\mu 3}$ events.

\begin{figure}
\centerline{\epsfysize 2.5 in \epsfbox{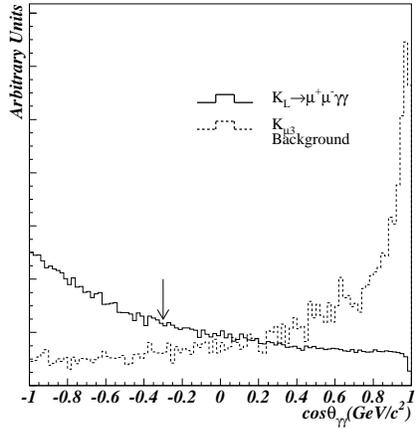}}   
\caption[]{The cosine of the angle between the two photons, from Monte 
Carlo simulations of
the signal (solid) and $K_{\mu 3}$ (dashed).  The requirement
of $\leq $ 0.3 is indicated by the arrow.}
\label{fig:costh}
\end{figure}

The cosine of the 
angle between the two photons in the kaon rest frame, 
$\cos\theta_{\gamma\gamma}$, was
also used to  reject $K_{\mu3}$ decays.  The distribution of
$\cos\theta_{\gamma\gamma}$ for the signal peaks at -1 corresponding
to anti-collinear emission of the two photons.  The 
$K_{\mu 3}$ background, which
has two accidental clusters identified
as photons, displays no such correlation.  Figure~\ref{fig:costh}
shows the expected $\cos\theta_{\gamma\gamma}$ distribution for signal and
$K_{\mu 3}$ background.  We required
 $\cos\theta_{\gamma\gamma}$  to be  -0.3 or less.
This cut rejected 85.3\% of the remaining $K_{\mu 3}$ events.

We also required the transverse shower 
shape for the photon clusters
to be consistent
with that expected from an electromagnetic process.  The 
$\chi^2$ of  the spatial distribution of
energy deposited in the calorimeter was used to identify clusters
as photons.  This cut reduced the remaining backgrounds due to accidental
energy by a factor of 4.5
while retaining 98.8\% of the signal events.

\begin{figure}
\centerline{\epsfysize 2.7 in \epsfbox{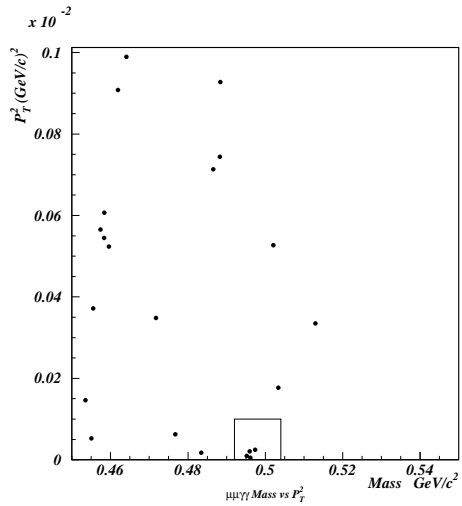}}   
\caption[]{$m$ vs. $P_\perp^2$ for the
events that passed all other cuts. The box is drawn around
the signal region, which contains 4 events.}
\label{fig:mass2d}
\end{figure}

In order to estimate the amount of background in the signal region,
we simulated all the leading sources of background.  
Our simulation incorporated 
both charged pion decay in flight and punch-through
the filter steel. 
The punch-through rate was a function of $\pi^{\pm}$ momentum,
determined by a $K_L\rightarrow \pi e \nu$ control sample.
The effect of accidental activity was simulated
by overlaying  Monte Carlo events with data 
from a random trigger that had a rate 
proportional to  the beam intensity.
The estimated 
background level  is detailed in Table~\ref{tab:backgrounds}.
A total of $0.155 \pm 0.081$ background events 
are expected within the signal region.
This region is defined by the
invariant mass of the 
$\mu^+\mu^-\gamma\gamma$ ($m$),
and 
square of the transverse momentum of the 
four particles with respect to a line connecting the target to
the decay vertex ( $P_\perp^2$ ) of the four particles in the
final state: 
$492 {\rm ~MeV/c}^2  <m < 504 {\rm ~MeV/c}^2$, 
and $P_\perp^2\leq 100 {\rm ~(MeV/c)}^2$.
After all the cuts 
we observed four events in the signal region.  
Figure~\ref{fig:mass2d} shows the $m$ vs. the
$P_\perp^2$ for events with all  but these cuts.  
A linear extrapolation of the high $P_\perp^2$ data in this figure yields
a background estimate of 0.25 $\pm$ 0.10 events, consistent with the 
expectation from Monte Carlo studies.  
To further test the background estimate with higher statistics we removed
the cluster shape $\chi^2$ cut and verified that the data matched the
prediction in $m$ and $P_\perp^2$ side bands.
The probability of observing four events in the signal region 
due to  fluctuation of the background  is
$2.1\times 10^{-5}$, corresponding to a 4.2 $\sigma$ fluctuation of
the estimated background.  
The branching ratio for 
$K_L\rightarrow\mu^+\mu^-\gamma\gamma$ was calculated
by normalizing the four signal events  to  a sample of 
$K_L\rightarrow \pi^+\pi^-\pi^0$ events, collected with the
prescaled normalization trigger. 
For the normalization events $m_{\gamma\gamma}$
was required to be within 3 MeV/$c^2$ of $m_{\pi^0}$,
and the ${\mathrm R}^{\pi\pi}_{\parallel}$ 
and muon counter hit requirements were not enforced.
The 
acceptance of these events was calculated to be 8.1\% via Monte Carlo.
We determined that $(2.68 \pm 0.04) \times 10^{11}$ $K_L$ 
within an energy range of 20 to 220 GeV decayed 
between 
90 and 160 meters from the target.  
The acceptance of the signal was (0.14 $\pm$ 0.01)\%,
so 
${\mathcal B}$($K_L\rightarrow \mu^+\mu^- \gamma \gamma$) $ =
(10.4^{+7.5}_{-5.9}  {\rm ~(stat)} )\times 10^{-9}$ with 
$m_{\gamma\gamma} \geq 1 {\rm ~MeV/c}^2$ which was the cutoff we used
in generating the Monte Carlo events.

We have  calculated the
branching ratio for this $K_L$ Dalitz decay  by 
performing a numerical
integration of the tree-level ($\mathcal O(\alpha)$)
$K_L\rightarrow \mu\mu\gamma\gamma$ matrix element
with an $m_{\gamma\gamma}\geq 1 {\rm ~MeV/}c^2$ cutoff. 
We performed a  similar integration of the 
$K_L\rightarrow\mu\mu\gamma$ matrix element, which 
included contributions due to virtual photon loops and emission
of soft bremsstahlung photons.  
Both integrations assumed unit form factors.   The
ratio of partial widths is 2.789\%.   
Multiplying this ratio with the  
measured value for 
${\mathcal B}(K_L\rightarrow\mu\mu\gamma)=(3.26 \pm 0.28)\times 10^{-7}$
~\cite{ref:mmg}
yields 
${\mathcal B}(K_L\rightarrow\mu\mu\gamma\gamma)=(9.1\pm 0.8)
\times 10^{-9}$.

The four-body phase space for $K_L\rightarrow \mu^+\mu^-\gamma\gamma$
can be parametrized by  five variables, 
as in reference \cite{greenlee}. 

\begin{figure}
\centerline{\epsfysize 3.2 in \epsfbox{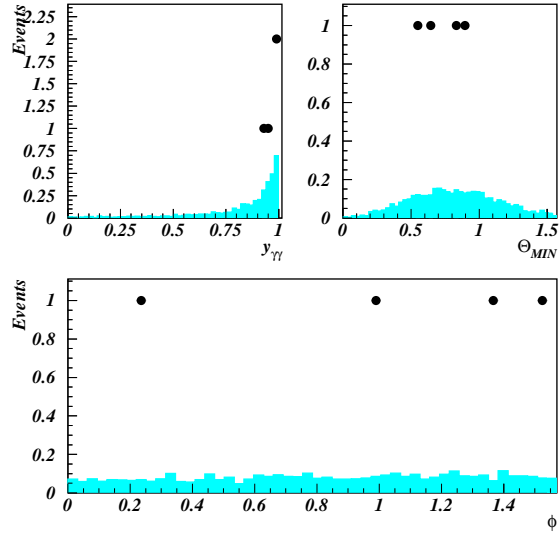}}   
\caption[]{The distributions for data (dots) and 
signal Monte Carlo (shaded) for the 
variables $y_\gamma$, $\Theta_{\rm MIN}$, and $\phi$
as defined in the text.}
\label{fig:greenlee}
\end{figure}

Figure~\ref{fig:greenlee} shows the distribution  of
the energy asymmetry of the photon pair ($y_\gamma$),  
the angle between the normals
to the planes containing the $\mu^+\mu^-$ and $\gamma\gamma$ in the center
of mass ($\phi$), and the minimum angle from any muon to any photon
($\Theta_{\rm MIN}$).  The distribution of these kinematic variables
for the four signal events is consistent with expectations.

We examined several possible 
sources of systematic uncertainty in the measurement.
The largest effects were due to a possible miscalibration of the 
calorimeter resulting in a mismeasurement of the photon energies,
and particle identification.
If we conservatively assume
a 0.7\% miscalibration of the calorimeter we obtain a 5.11\% systematic
error.  The  uncertainty due to muon identification was 
determined to be 4.2\% by comparing the $K_L$ flux with that
obtained by using $K_{\mu 3}$ decays.
The uncertainty in the $K_L\rightarrow
\pi^+\pi^-\pi^0$ branching ratio is 1.59\%.  
Adding  these and other smaller contributions
detailed in Table~\ref{tbl:systs} in quadrature
we assigned a total systematic
uncertainty of 6.95\% to the branching ratio measurement.  

In summary we have determined 
the branching ratio to be
${\mathcal B}$($K_L\rightarrow \mu^+\mu^- \gamma \gamma$) $ =
(10.4^{+7.5}_{-5.9}  {\rm ~(stat)} \pm 0.7 {\rm ~(sys)})\times 10^{-9}$ with 
$m_{\gamma\gamma} \geq 1 {\rm ~MeV/c}^2$.  Defining the acceptance with
a 10 MeV 
infrared cutoff for photon energies in the kaon frame
($E_{\gamma}^*$), our result is 
${\mathcal B}$$(K_L\rightarrow\mu^+\mu^-\gamma\gamma ; 
E_{\gamma}^* \geq 10 {\rm ~MeV}) 
 = (1.42 ^{+1.0}_{-0.8} {\rm ~(stat)} \pm 0.10 {\rm ~(sys)})\times 10^{-9}$.
This is the first observation of this decay
and is consistent with theoretical predictions.

We gratefully acknowledge the support and effort of the Fermilab
staff and the technical staffs of the participating institutions for
their vital contributions.  This work was supported in part by the U.S. 
DOE, The National Science Foundation and The Ministry of
Education and Science of Japan. 
In addition, A.R.B., E.B. and S.V.S. 
acknowledge support from the NYI program of the NSF; A.R.B. and E.B. from 
the Alfred P. Sloan Foundation; E.B. from the OJI program of the DOE; 
K.H., T.N., K.S., 
and M.S. from the Japan Society for the Promotion of
Science.  


\begin{table}[b]
\caption[]{The various backgrounds to 
$K_L\rightarrow\mu^+\mu^-\gamma\gamma$.  $\pi^{\pm}$ can can be mistaken
for $\mu^{\pm}$ due to decay (D) or punch-through (P).  Accidental clusters
in the calorimeter identified as photons are designated $\gamma_{acc}$.}
\vspace*{0.2in}
\begin{tabular}{|l|c|c|}
\multicolumn{1}{|c|}{\bf Decay}  & 
{\bf Cause of  $\mu$ misid}  & {\bf Events expected} \\
\hline
$K_L\rightarrow \mu^+\mu^-\gamma\gamma_{acc}$  & & 0.093 $\pm$ 0.036 \\
$K_L\rightarrow \pi^+\pi^-\pi^0$ &  DD & $<$ 0.056 \\
$K_L\rightarrow \pi^+\pi^-\pi^0$ & DP  & $<$ 0.011 \\
$K_L\rightarrow \pi^+\pi^-\pi^0$ & PP  & $<$ 0.011 \\
$K_L\rightarrow \pi^{\pm}\mu^{\mp}\nu + 2\gamma_{acc}$ & D  & 0.030$\pm$0.030 \\
$K_L\rightarrow \pi^{\pm}\mu^{\mp}\nu + 2\gamma_{acc}$ & P  & 0.032$\pm$0.032 \\
$K_L\rightarrow \pi^{\pm}\pi^0\mu^{\mp}\nu$ & D  & $<$0.005 \\
$K_L\rightarrow \pi^{\pm}\pi^0\mu^{\mp}\nu$ & P  & $<$0.004 \\
\hline
\multicolumn{2}{|c|}{\bf Total} & 0.155 $\pm$ 0.081 \\
\end{tabular}
\label{tab:backgrounds}
\end{table}
\begin{table}[b]
\caption{Systematic and statistical sources of uncertainty.  Sources
marked with (*) contribute to uncertainty in both the 
$K_L$ flux and the
acceptance for 
$K_L\rightarrow \mu\mu\gamma\gamma$ 
relative to the acceptance for 
\klpmz; other
sources contribute only to the acceptance ratio.}
\label{tbl:systs}
\begin{tabular}{lc}
\multicolumn{1}{c}{Source}           & Relative Uncertainty \\
\tableline
${\mathcal B}${\klpmz}               & 1.59\% (*)           \\
Data statistics for \klpmz\          & 0.16\% (*)           \\
Simulation statistics                & 0.22\% (*)           \\
Calorimeter scale and resolution     & 5.11\%               \\
Spectrometer scale and resolution    & 0.98\%               \\
Muon identification                  & 4.20\%               \\
Signal trigger requirements          & 0.80\%               \\
Vertex quality requirement           & 0.24\%               \\
Spectrometer wire inefficiency       & 0.37\%               \\
\tableline
\multicolumn{1}{c}{Total}            & 6.95\%               \\
\end{tabular}
\end{table}

 
\end{document}